\documentclass[preprint]{aastex}

\shorttitle{Study of V616 Mon}
\shortauthors{Gelino, Harrison, \& Orosz}

\begin{document}

\title{A Multi-Wavelength, Multi-Epoch Study of the Soft X-Ray Transient 
Prototype, V616 Mon (A0620-00)\altaffilmark{1}}

\author{Dawn M. Gelino,\altaffilmark{2} Thomas E.
Harrison\altaffilmark{3} \email{dleeber@nmsu.edu, tharriso@nmsu.edu}}
\affil{Department of Astronomy, New Mexico State University, Las
Cruces, NM 88003} \author{Jerome A. Orosz
\email{J.A.Orosz@astro.uu.nl}} \affil{Astronomical Institute, Utrecht
University, PO Box 80000, NL-3508 TA Utrecht, The Netherlands}

\altaffiltext{1}{This work was partly based on observations obtained
with the Apache Point Observatory 3.5-meter telescope, which is owned
and operated by the Astrophysical Research Consortium.}

\altaffiltext{2}{Visiting Astronomer, Kitt Peak National Observatory, 
National Optical Astronomy Observatory, which is operated by the Association 
of Universities for Research in Astronomy, Inc. (AURA) under cooperative 
agreement with the National Science Foundation.}

\altaffiltext{3}{Visiting Astronomer, Cerro Tololo Inter-American Observatory, 
National Optical Astronomy Observatory, which is operated by the Association 
of Universities for Research in Astronomy, Inc. (AURA) under cooperative 
agreement with the National Science Foundation.}

\begin{abstract}
We have obtained optical and infrared photometry of the soft x-ray
transient prototype V616 Mon (A0620-00).  From this photometry, we
find a spectral type of K4 for the secondary star in the system, which
is consistent with spectroscopic observations.  We present $J$-, $H$-, and
$K$- band light curves modeled with WD98 and ELC.  Combining detailed,
independently run models for ellipsoidal variations due to a spotted, 
non-spherical secondary star, and the observed ultraviolet to infrared 
spectral energy distribution of the system, we show that the most likely 
value for the orbital inclination is $40.75\pm 3^{\circ}$.  This 
inclination angle implies a primary black hole mass of 
$11.0\pm 1.9\,M_{\odot}$.

\end{abstract}

\keywords{binaries: close
$-$ stars: black holes
$-$ stars: individual (V616 Mon)
$-$ stars: low mass
$-$ stars: variables: other}

\section{Introduction}

Soft X-Ray Transients (SXTs) are a family of low mass X-ray binaries that
display infrequent, but large and abrupt X-ray and optical outbursts 
that are believed to be the result of a sudden, dramatic increase in the 
mass accretion rate onto the compact object.  In most cases the compact
object is a black hole and the companion star is a low-mass K- or M-type
dwarf (see Charles 2001 for a recent review). During their long periods of 
quiescence, SXTs are very faint at X-ray and optical wavelengths, however, 
in this state, the secondary stars can dominate the system luminosity and 
allow us to derive the SXT parameters. 

The SXT prototype, V616 Mon (=A0620-00) ($\alpha_{2000}$ = 06${\rm
^h}$22${\rm ^m}$44${\rm ^s}$.4, $\delta_{2000}$ = -00$^{\rm
o}$20$\arcmin$45$\arcsec$) was discovered by the Ariel-5 satellite on
1975 August 3 \citep{elv75}, while the optical counterpart to the
X-ray source was identified by \citet{bol75}.  By studying the
optical spectrum of the system, \citet{oke77} found that one component of 
the light came from a K5 - K7 dwarf, and another from a suspected
accretion disk.  Since V616 Mon has faded from outburst, it has been
studied by \citet{mcc86}, \citet[hereafter HRHSA]{has93},
\citet[hereafter MRW]{mar94}, \citet[hereafter SNC]{sha94},
\citet[hereafter FR]{fro01}, and references therein.  These authors
have found an orbital period of 7.75 hours, a secondary star spectral
type of K3V - K4V, and a secondary star radial velocity semi-amplitude
of 433$\pm$3 km s$^{-1}$.  V616 Mon therefore has an implied 
optical mass function, which is the minimum mass of the compact primary 
object, of 2.72$\pm$0.06 M$_{\odot}$ (MRW).

\citet{mcc00} report the X-ray luminosity of V616 Mon in quiescence is
$L_{X}\approx 6 \times 10^{30}$ erg s$^{-1}$, and exhibits a very low 
accretion rate ($\dot{M}_d \sim 10^{-10}$ $M_{\odot}$ yr$^{-1}$).   
Therefore, the current quiescent light curves should be primarily 
governed by the light from the secondary star.  Since the secondary star 
fills its Roche lobe, the amount of surface area seen by an observer on 
Earth changes as the star orbits the compact object.  This changing 
line-of-sight surface area corresponds to a changing apparent brightness, 
giving rise to the so-called ellipsoidal variations. The amplitude of 
these variations is determined by the orbital inclination angle and mass
ratio of the system.  By combining the orbital inclination angle with the 
observed mass function, the SXT system parameters can be determined.  
In an effort to quell the debate over the inclination angle of the SXT 
prototype, we have obtained new optical and multi-epoch infrared 
photometry of V616 Mon.  We have used two different light curve modeling 
programs to interpret the infrared ellipsoidal varaitions observed over a 
22.5 month interval.  We have previously used a similar technique to 
model GU Mus \citep[hereafter Paper 1]{gel01}, an SXT that closely 
resembles V616 Mon.

Most previous attempts to derive the orbital parameters for V616 Mon
have used optical data.  If one is searching for the purest ellipsoidal 
variations, one should observe at a wavelength where the secondary star 
provides the majority of the systemic luminosity.  In the optical, the 
accretion disk and hot spot can contaminate, if not dominate, the system 
luminosity.  Even in quiescence, there could be a modest amount of dilution 
of the optical light curve by the accretion disk, hot spot, and/or 
(although somewhat less likely) heating of the secondary star due to the 
possibility of weak and variable accretion \citep{sha97}.  On the other 
hand, in the infrared, the late type secondary stars can dominate the 
quiescent binary's luminosity.  This reduces the uncertainties in modeling 
the system.  For V616 Mon, MRW find that the K star contributes 83$\pm$4\%
in the blue, and 94$\pm$3\% of the contimuum flux near H$\alpha$. This
percentage should be even higher at longer wavelengths.  Therefore,
observations in the infrared will reveal more genuine ellipsoidal
variations than observations in the optical regime.

SNC and FR both observed V616 Mon in the infrared.  SNC obtained a
$K$- band light curve, while FR obtained $H$- band light curves as
well as much more sparsely sampled $J$- and $K$- band curves.  Neither
study could accurately determine the amount of contamination present
from other sources of light in the system.  Because of this, the
orbital inclination angle determined from FR spanned nearly the entire range
of published results ($38^{\circ}\le i \le 75^{\circ}$).  In order
to determine if contamination is present, multi-wavelength
observations are needed.

Currently, the best way to obtain the orbital inclination angle in 
non-eclipsing binary systems is to model the ellipsoidal light curves.
Previously published 
inclination angles for V616 Mon range from $31^{\circ}$ (lower limit 
from SNC) all the way to $75^{\circ}$ (upper limit from FR).  These angles
correspond to primary masses of $28.8\,M_{\odot}$ to $3.3\,M_{\odot}$,
respectively.  The large range in derived inclination angles (and
masses) appears to be due in part to the changing shape of the
observed light curves over time.  In 1986 December and 1987 January,
HRHSA saw evidence for a ``grazing eclipse of the Roche lobe-filling
star by the accretion disk'', but when MRW observed the SXT in 1991
December and 1992 January, this feature was no longer present.  If the
eclipse was real, it would indeed suggest a high orbital inclination
angle.  However, no other published data on V616 Mon has shown
evidence for such events.  The most recently published data set on
V616 Mon was taken by FR in 1995 and 1996.  The data showed that as
the $H$- band light curve shape exhibited slight changes over the one 
year baseline of their observations, the mean $H$ magnitude remained constant.

We have undertaken a program to attempt to determine more precise
orbital inclination angles and primary masses for five SXTs.  We have
observed these SXTs in the infrared, and have already presented
results on GU Mus (Paper 1).  We now present our results for the
prototype SXT, V616 Mon, based on a two year baseline of multi-wavelength 
optical and infrared observations.  We model these data with both WD98, 
as done in Paper 1, and independently with the ELC code \citep{oro00}. 
In section 2, we describe our observations and data reduction process.  
In section 3, we present our infrared photometric light curves and discuss 
the possibility of long and short term variablility. Section 4 describes
our choices for the relevant input parameters for WD98 and ELC. We
present the resulting models at $J$, $H$, and $K$, and discuss the
possible scenarios to explain the changing shape of the long-term
light curves. Finally, in Section 5 we discuss the implications of the 
models, and compare our results to those previously published.

\section{Observations \& Data Reduction}

We obtained infrared photometry of V616 Mon on three seperate epochs
over a 22.5 month interval.  We first observed the SXT prototype using
GRIM II\footnote{See http://www.apo.nmsu.edu/Instruments/GRIM2/} on
the Astrophysical Research Consortium 3.5-m telescope at Apache Point
Observatory on 1999 February 25.  We then used IRIM\footnote{See
http://www.noao.edu/kpno/manuals/irim/} on the 2.1-m telescope at the
Kitt Peak National Observatory on 2000 February 12 and 16, and had our
third and final observing run on the same telescope using
SQIID\footnote{See http://www.noao.edu/kpno/sqiid/sqiidmanual.html} on
2000 December 8, 9, and 11.

On 1999 February 25, V616 Mon was observed from 3:43 to 7:41 UT with
the camera at the f/5 plate scale (0$\arcsec$.473 pixel$^{-1}$).  With
an orbital period of about 7.75 hours, this first observing session
covered just over half of an orbital period.  Photometric data were
obtained in the GRIM II $J$ ($\lambda_c=1.265 \micron$) filter. Our
observing sequence consisted of two 15 second images at one position,
a beam switch, and two additional 15 second images.  All of the data
were linearized with an IRAF linearization task written by A. Watson
(1996, Private Communication).  After averaging the two images at one
position, we subtracted them from the average of the two images at the
other position.  These sky, and bias-subtracted images were then flat
fielded with a sky flat using the usual IRAF packages.

V616 Mon was observed again on 2000 February 12 from 4:16 to 6:41 UT
and on February 16 from 1:45 to 8:06 UT with the camera at the f/15
focus (1$\arcsec$.09 pixel$^{-1}$). This second observing run
effectively covered one orbital period.  Data were obtained in the
IRIM $J$ ($\lambda_c=1.24 \micron$), $H$ ($\lambda_c=1.65 \micron$), 
and $K'$ ($\lambda_c=2.16 \micron$) filters.  Our observing sequence 
was the same as that for the GRIM II data.  We began with the $K'$ 
filter, refocussed and switched to the $H$ filter, and repeated the 
procedure for the $J$ filter before returning back to the $K'$ filter 
and refocussing again.  Each individual $J$ image consisted of 1 frame 
of 30 seconds, the corresponding $H$ images consisted of 6 coadded 
frames of 8 seconds each, while each $K'$ image consisted of 10 coadded 
frames of 4 seconds each.  We processed the images as above, this time 
linearizing the data using the IRLINCOR package in IRAF with the 
coefficients supplied in the IRIM User's Manual.
	
We observed V616 Mon for the final time on 2000 December 8 from 4:59
to 12:29 UT, on December 9 from 4:36 to 12:34 UT, and on December 11
from 6:42 to 12:18 UT, all with the camera at the f/15 focus
(0$\arcsec$.69 pixel$^{-1}$).  This final observing run covered 2.72
orbits.  Data were obtained in the SQIID $J$ ($\lambda_c=1.267 \micron$), 
$H$ ($\lambda_c=1.672 \micron$), and $K$ ($\lambda_c=2.224 \micron$) 
filters.  This time, we observed with an ABBA sequence.  SQIID stands 
for ``Simultaneous Quad Infrared Imaging Device'', and takes images in 
4 filters simultaneously.  Due to this, the number of coadds and exposure 
times for each filter had to be identical.  Each individual image consisted 
of 8 coadded frames of 7 seconds each.  We chose this exposure time to 
stay safely below the non-linear regime of the chip since detailed 
linearization curves do not yet exist for this newly commissioned 
instrument.  Dome flats were taken due to the long readout time per 
exposure.  We again processed the images as above with IRAF.

For each of the three data sets, we performed aperture photometry on
V616 Mon and the same five nearby field stars.  Using the IRAF PHOT
package, a differential light curve in each wavelength regime was
generated with each point being the average of four beam switched
images.  For the GRIM II and SQIID data sets, our differential
photometric results show that over the course of our observations, the
comparison stars did not vary more than expected from photon
statistics.  Unfortunately, while the S/N of the IRIM data set was
high ($\sim 1 \%$ relative photometry), these data showed considerable
scatter due to very good seeing which resulted in the undersampling of
the PSF.

Optical observations of V616 Mon were taken with the Cassegrain Focus
CCD Imager\footnote{See http://www.ctio.noao.edu/cfccd/cfccd.html} on
the 0.9 meter telescope at CTIO on 2001 March 15 at 00:50 UT.  Data
were obtained in the $B$, $V$, $R$, and $I$ bandpasses for the purpose
of determining the quiescent optical colors of the system.  The 180
second exposures (300 seconds for the $B$ band) were zero corrected
and flat fielded before aperture photometry was performed.  Standards
were also observed to transform these data to the system of Landolt.
The observed optical and infrared colors of V616 Mon can be found in
Table 1, along with previously published optical colors for this
system.

\section{Long and Short Term Variability of the Light Curve}

Since the 1975 outburst of V616 Mon, several variations of light curve
shape have been observed.  A few years after the system's transition
into its quiescent state, \citet{mcc86} obtained $''B+V''$- and $I$- band
light curves.  In both of these light curves, the deeper minimum
coincided with the inferior conjunction of the secondary star.  When
HRHSA observed the system in 1986 and 1987, the deeper minimum instead
corresponded to the inferior conjunction of the primary object.  In
1990, \citet{bar91} obtained a light curve with minima consistant with
HRHSA's, but their brightest maximum was observed to be shifted by 0.5
in phase.  Similar changes were found by \citet{lei98} when they
studied the $R$- band light curve of V616 Mon over a seven year
period. They found that not only did the shape of the light curve
change over time, but also the mean brightness of the system seemed to
vary on a time-scale of hundreds of days with a peak-to-peak amplitude 
of 0.3 mag.  More recently, FR saw evidence for a slightly changing $H$- 
band light curve shape over a one year baseline of observations.  However, 
they did not see any sharp dips or asymmetry reversals between the two 
light curve maxima. In addition, the mean $H$- band magnitude remained 
fairly constant over the course of their observations.

\begin{figure}
\figurenum{1}
\plotone{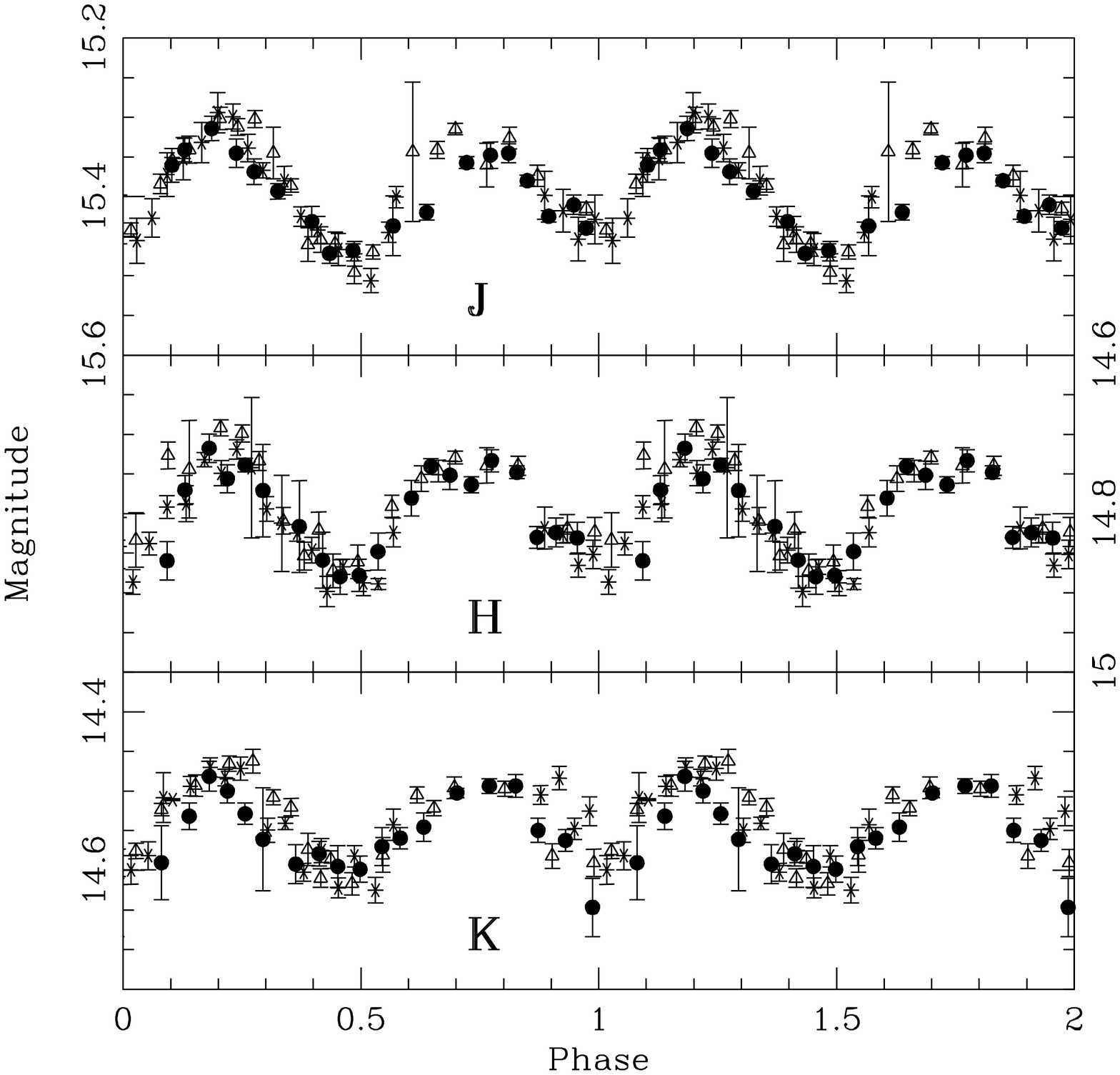}
\caption{SQIID $J$-, $H$-, and $K$- band data taken on 2000 December 8
(filled circles), 9 (open triangles), and 11 (asterisks), with the 2.1
meter telescope at Kitt Peak National Observatory.  The data are plotted 
over two phase cycles for clarity and are phased to the \citet{lei98} 
ephemeris shifted by 0.5 in phase so that phase 0.0 represents the inferior 
conjunction of the secondary star.  Error bars are 1-$\sigma$.  There are 
no significant night-to-night variations. \label{fig1}}
\end{figure}

\begin{figure}
\figurenum{2}
\plotone{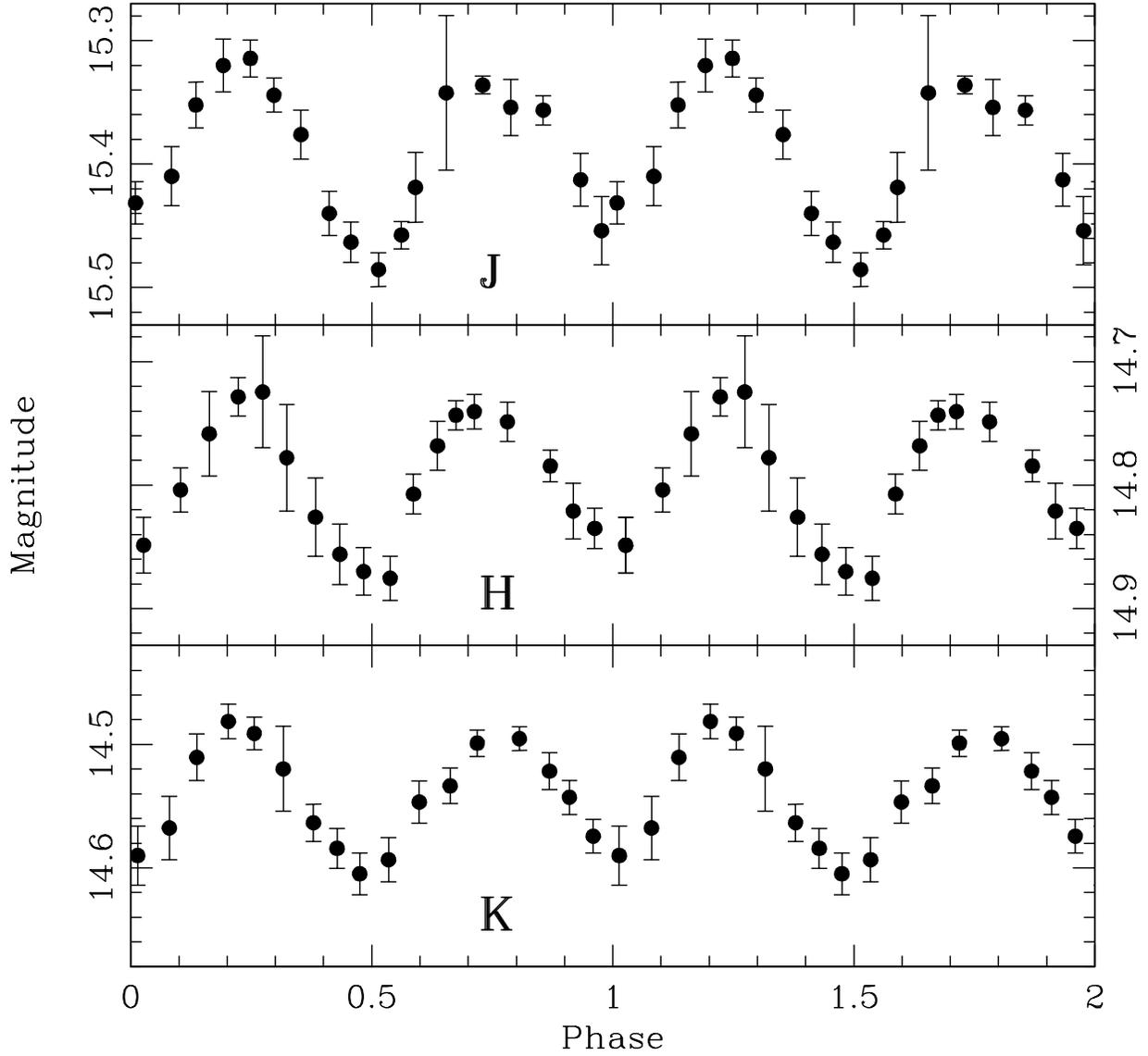}
\caption{Final 2000 December V616 Mon $J$-, $H$-, and $K$- band light
curves. Here, the data from Figure \ref{fig1} were binned in phase in order to
obtain one light curve per band. Error bars are 1-$\sigma$. \label{fig2}}
\end{figure}

With the history of V616 Mon's changing light curve shape, we were
interested in determining its current level of activity.  We observed
V616 Mon over a roughly two year period including three nights in
December 2000.  In Figure \ref{fig1}, we present the $J$-, $H$-, and $K$- 
band light curves of V616 Mon obtained on 2000 December 8, 9, and 11.  The
data were phased to the \citet{lei98} ephemeris, and then shifted by
0.5 in phase so phase 0.0 corresponds to the inferior conjunction of
the secondary star (where the secondary star is in front of the black
hole from the viewpoint of the observer).  Each symbol represents a
different night of observations.  There are no strong variations in
the amplitude or mean brightness of the SXT from night to night. This
indicates that the infrared light curves are stable over time periods
of days.  Since each night of observation revealed a light curve shape
consistent with the other two, we binned the points in a given phase
interval and combined the three nightly light curves into one for each
wavelength regime.  The results are presented in Figure \ref{fig2}. Our
infrared light curves most closely resemble the optical light curves
of HRHSA, with the brighter maximum occurring at phase 0.25 and
deeper minimum at phase 0.5.  If we compare our $H$- band light curve
to FR's, we find that the deeper minimum occurs at phase 0.5 in both
curves, but the brighter maximum differs by 0.5 in phase.

\begin{figure}
\figurenum{3}
\plotone{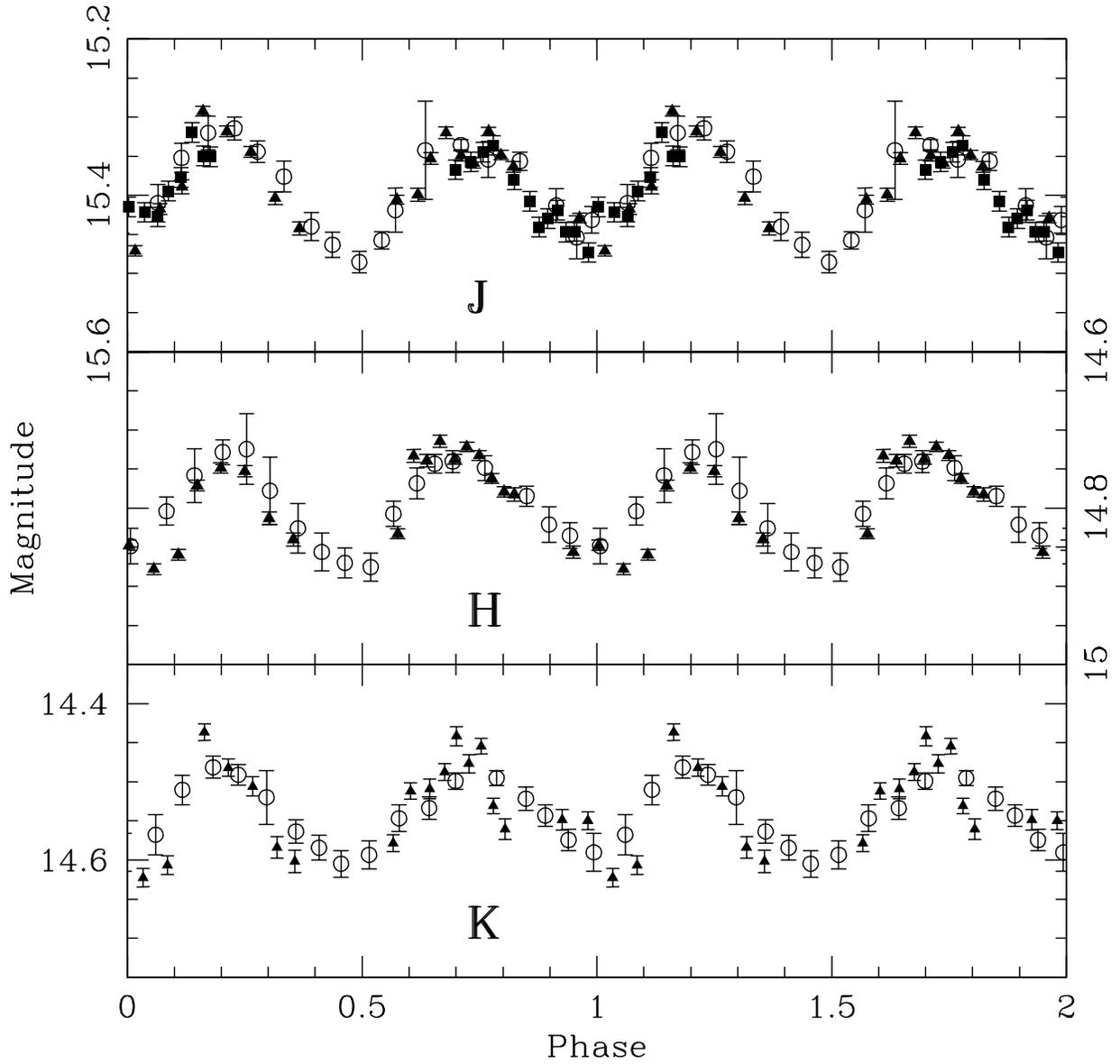}
\caption{Long-term infrared light curves of V616 Mon. The $J$- band
data span 22.5 months while the $H$- and $K$- band data span 10.5
months.  The open circles represent the SQIID data taken on 2000
December 8, 9, and 11 as plotted in Figure \ref{fig2}.  The filled triangles
represent the data taken with IRIM at the 2.1 m telescope at Kitt Peak
National Observatory on 2000 February 12 and 16.  The filled squares
in the $J$- band light curve represent data taken using GRIM II on the
Astrophysical Research Consortium 3.5 m telescope at Apache Point
Observatory on 1999 February 25.  There are no obvious long-term
variations in the shapes or mean magnitudes of the light curves. \label{fig3}}
\end{figure}

Since the shape of our 2000 December light curves differed from FR's
taken in 1995 and 1996 (the $H$- band magnitudes were consistent
within the errors), we utilized our almost two year baseline of
observations to investigate if the shape of the light curve had
changed over this time.  Figure \ref{fig3} presents the $J$-, $H$-, 
and $K$- band light curves phased as in Figure \ref{fig1}.  Unfortunately, 
while the S/N of the IRIM data set (filled triangles) was high ($\sim 1 \%$
photometry), these data showed considerable scatter due to very good
seeing which resulted in the undersampling of the PSF.  In any case,
both the shape and brightness of the observed light curves are
consistent over the period of observations.  Nonetheless, the evidence
for long-term changes in the shape of the quiescent light curves of
V616 Mon is interesting, and we will return to this subject in the next 
section.

\section{Modeling the Infrared Light Curves of V616 Mon}

We adopt as our light curves the binned $J$-, $H$-, and $K$-band light
curves from 2000 December (Figure \ref{fig2}).  In order to find the 
inclination of the binary, we conducted two more or less independent 
analyses of these data, which we describe below.

\subsection{Basic Models and Assumptions}

\subsubsection{Assumed Secondary Star Parameters}

For multi-wavelength light curve modeling, the most important input
parameters are based on the temperature of the secondary star.  In
1977, Oke detected a cool spectrum from the source and attributed it
to a K5V - K7V star.  HRHSA's analysis of V616 Mon favored a spectral type
of K3 - K4.  \citet{sha99} compared their observed $K$- band spectrum
of V616 Mon to the spectra of comparison stars of spectral types K0V,
K3V, K5V, and K7V, and using $\chi^2$ tests, determined a best fit
spectral type of K3V for the secondary star.  Here, we examine
multi-wavelength photometry to estimate the secondary star's spectral
type.

Table 1 presents the post-outburst infrared and optical colors of V616
Mon.  The mean colors are consistent with those of a K4V star reddened by
$A_{V} = 1.21$ mag \citep[$E(B-V)=0.39$]{wu76, sea79}.  In Figure \ref{fig4}
we compare the dereddened observed spectral energy distribution (SED)
of V616 Mon to a K4V star \citep{bes88, cou76, joh66, mik82}.  
Except at $B$ and $J$, the two sets of data (normalized at $H$) are 
consistent within the errors.  Based on the photometry, there is evidence 
for excess light at $B$ (22\%) and a possible small excess at $J$.  
This $B$-band excess is consistent with that found by MRW, and helps us 
to constrain the amount of contamination by the accretion disk and/or 
hot-spot. Based on the mean observed colors of V616 Mon, 
and its resulting spectral energy distribution, we adopt a spectral type 
of K4 for the secondary star.  We adopt as the corresponding temperature 
a value of $T_{\rm eff}=4600$~K \citep{joh66}.

Given that the photomtery and optical/infrared spectroscopy agree, the 
simplest assumption is that the K4 secondary star dominates the $VRIJHK$ 
luminosity of the system.  We can, however, envision scenarios (some of 
which are highly tuned) where a K-type SED might be mimicked by accretion 
processes.  Without accurate models for the spectra of quiescent SXT 
accretion disks, we can not further speculate on the exact level of 
contamination associated with these processes and therefore assume that 
nearly all of the infrared light in the system comes from the secondary star.

The seconday stars in systems such as SXTs can have masses and radii that
do not correspond to accepted values for main sequence stars.  In
addition, some of these stars have been found to have parameters
consistent with those of evolved stars (e.g. Paper 1).  Mass estimates
range from that of a zero-age main sequence star (ZAMS) to those of a
less massive subgiant \citep[MRW]{ibe97}.  Fortunately, we do not need
a value for the mass of the secondary star as long as we have a good
measurement for the mass ratio, $q$.  Unlike GU Mus (Paper 1), V616
Mon has a well determined mass ratio. MRW measured the mean rotational 
velocity of the secondary star. Their value of $v \sin i=83\pm 5$ 
km s$^{-1}$ implies a mass ratio of $q\equiv M_2/M_1=0.067\pm 0.01$, 
assuming the secondary star is rotating synchronously with the orbit.  
It turns out that the derived orbital inclination angle is
relatively insensitive to the value of $q$; however, $q$ will be
important in the determination of the primary object's mass once the
orbital inclination angle is known.

\subsubsection{WD98 Model Setup}

The first of the models used was WD98, the newest version of the 
Wilson-Devinney light curve program \citep[J. Kallrath 1999, private 
communication; R. E. Wilson 1999, private communication]{wil98}.  See 
Paper 1 for references and a basic description of the code. We ran WD98 
in a mode for a semi-detached binary with the secondary star 
automatically filling its Roche lobe (Mode = 5), and the primary having 
such a large gravitational potential, that it essentially has zero radius.
The secondary star atmosphere was determined from solar metallicity Kurucz 
models.  See Paper 1 for a discussion of the parameters such as limb 
darkening, gravity darkening, and bolometric albedo.  The most important
wavelength-independent input values to WD98 are listed in Table 2, and the
wavelength-dependent parameters in Table 3 (units are shown where
appropriate).  Since any realistic hot spot or disk contamination for 
this system will have a minimal heating effect on the secondary star 
(HRHSA; FR), we have used normal, non-irradiated, limb darkening 
coefficients in our models.  In addition, we adopt the gravity darkening 
exponent found by \citet{cla00} for a K4 star ($T_{\rm eff} = 4600$ K):
$\beta_1 = 0.40$ (Table 2).

We did not make use of the optimizer program supplied with WD98.  It
proved to be much more convenient to simply compute large grids of models
and use a simple program to compute $\chi^2$ values of the fits.  
The mass ratio was usually fixed at its spectroscopic value.
In order to quantify the sensitivity of the models to the variations
of the input parameters, we ran models with inputs covering a wide
range of parameter space as in Paper 1.  We found that the best fit 
orbital inclination angle did not change significantly as we varied the
temperature of the secondary star by one spectral type. Varying $q$
from 0.057 to 0.077 (the $1\sigma$ error bars from MRW), also had no 
effect on our final orbital inclination angle.

\subsubsection{ELC Model Setup}

The second code we used was the ELC program, which is fully described in 
\citet{oro00}.  In its ``black body'' mode ELC is essentially the same 
as modes 2, 4, 5, and 6 of WD98.  ELC has at least two features with no 
direct analog in WD98.  The first is the (optional) inclusion of a 
flared accretion disk around ``star \#2''.  The second is the way in 
which model atmosphere intensities are included in the model. WD98 
tabulates the ratios of solar metallicity model atmosphere intensities 
to black body intensities for surface normals (i.e.\ a viewing angle of 
$\mu=1$).  The intensities for other viewing angles are computed using 
the specified limb darkening law (a single limb darkening law is used for 
the entire star). ELC tabulates model solar metallicity atmosphere specific 
intensities \citep{hau99a,hau99b} as a function of three parameters, namely 
the effective temperature, the surface gravity, and the viewing angle $\mu$.  
Hence no parameterized limb darkening law is needed.  For stars with large 
surface gravities ($\log g > 3.5$) the systematic differences between ELC 
and WD98 are quite small (less than $\approx 1\%$).  However, for cool 
giants, the differences in the light curves can be dramatic \citep{oro00}.

ELC currently has four different optimizer routines with various levels
of sophistication.  They are the ``grid search'' method and a
Levengerg-Marquardt routine adapted from \citet{bev69}, a routine
which uses the downhill simplex method (essentially the ``amoeba''
routine from Press et al.\ 1992), and a genetic algorithm based on
the ``pikaia'' routine given in \citet{cha95}.  The genetic routine
proved to be by far the most useful routine here. The three basic free 
parameters of the model are the mass ratio, the inclination, and the 
orbital separation.  The binary observables that were modeled were the 
three infrared light curves, a simulated velocity curve with 
$K_2=433\pm 3$ km s$^{-1}$ (MRW), and MRW's value of the mean rotational 
velocity of the secondary star ($83\pm 5$ km s$^{-1}$).  
The optimization routines attempt to find a parameter set so that the
total $\chi^2$ is minimized ($\chi^2_{\rm total}=\chi^2_{\rm light}
+\chi^2_{\rm velocity}+\chi^2_{\rm rotation}$).

\subsection{Simple Ellipsoidal Models of the Secondary Star} 
                                                         
We first ran models assuming the secondary star was the sole source of
light.  Aside from limb- and gravity- darkening effects, the secondary 
star was assumed to have a uniform surface brightness.  

The WD98 models were run for a range of inclination angles, using the 
input parameters listed in Tables 2 and 3. The best fit was determined 
based on $\chi^2$ tests.  In Figure \ref{fig5}, we present our 2000 December 
V616 Mon $J$-, $H$-, and $K$- band observations and the corresponding 
WD98 model (solid line) for the best fit value of the orbital inclination 
angle $i = 40^{\circ}$.  

Three iterations of the ELC grid search program were used to fit the data 
and the results are shown as the dashed line in Figure \ref{fig5}.  The fitted 
parameters are $i=40.03^{\circ}$, $q=0.0604$, and $a=4.543\,R_{\odot}$.  
The WD98 and ELC models are nearly indistinguishable.

It is clear that the infrared variations we observed are not purely
ellipsoidal.  All three light curves reveal a small difference between
the two maxima, as well as a small deviation ($\sim$ 3\% at $K$) between 
phases 0.0 and 0.25.  There are several possible explanations that could 
account for this light curve shape, and we disuss them in the following 
sections.

\subsection{Possible ``Contamination'' Scenarios}

There are two main effects that could result in a phase dependent
deviation from ellipsoidal variations.  These possibilities are
direct light from a hot spot on the edge of the accretion disk, and spots
on the secondary star, either cool (dark) or hot (bright) spots.  
The long-term changing shape of the light curves discussed in Section 3 
may be explained if the level of heating or the positions of these spots 
vary over time.  We will consider each of these possibilities in turn.

\subsubsection{A Migrating Accretion Disk Hot-Spot}

In 1991, MRW ``imaged'' V616 Mon through doppler tomography, and found
evidence for a hot spot on the accretion disk.  The hot spot is 
presumably formed when the accretion stream from the secondary star impacts 
the accretion disk.  The accretion stream has a predicted path, therefore,
in order for the hot spot to migrate around the edge of the disk, the
disk size must change.  There have been suggestions that the accretion
disk in V616 Mon has changed in size over the past two and a half
decades.  For example, based on the change in the peak separation of 
observed H$\alpha$ lines between 1991 and 1993, and its corresponding 
change in the disk velocity ($\Delta$ V = 100 km s$^{-1}$), \citet{oro94} 
suggested 
that the accretion disk had contracted.  Thus, the accretion disk may indeed 
change in size causing the hot spot to migrate. The hot spot could introduce 
a phase dependent source of light if the accretion disk was optically thick.  
If the spot were more or less on the disk rim, then its viewing angle would 
change considerably over the orbit, giving rise to a phase-dependent 
modulation.

We investigated several spotted disk models using ELC and its genetic code.
There are several extra parameters needed to describe the disk.  They
are the inner and outer radii, the opening angle of the outer edge,
the temperature at the inner edge, and the power-law exponent which
describes the temperature profile.  In addition, four parameters are
needed for a spot.  They are the azimuth angle, the angular size of the 
spot, its radial extent, and its ``temperature factor''.  About 30,000 
models using a wide range of input parameters were computed and fitted 
using the genetic code.  In the end we found what FR had found, namely 
that one can find models with a wide range of inclinations that give nearly 
equally good fits.  In our case the inclinations ranged from about 40 
degrees to 60 degrees with a change in the total $\chi^2$ of less than 1.
Fortunately, with the use of multi-color photometry and the observed 
spectral energy distribution, we are able to exclude a large number of 
these models.  These models generally predict that on the order of 50\% 
of the light in the infrared should come from the accretion disk.  The 
``disk fraction'' in the optical is predicted to be even larger. The 
ultra-violet to infrared SED of V616 Mon (Figure \ref{fig4}) limits the excess 
light from the accretion disk and hot-spot to 22\% at $B$, and $\le$3\% 
at all other visible and near-infrared wavelengths. Although we cannot 
rule out up to a $\approx 3$\% level of diluting infrared light from the 
accretion disk and/or hot spot, we can certainly rule out the large 
dilutions predicted by the spotted disk models.  Based on these arguments, 
we do not believe that direct light from the hot-spot is the cause of the
excess in the ellipsoidal variations.

\subsubsection{Star Spots}

Star spots are believed to be common features on many late-type stars.  
Their colors will not be drastically different from a K4V star, and 
therefore do not significantly alter the observed SED.  Spots on late-type 
stars typically vary on timescales of months to a few years \citep{bou88, 
vog75}. \citet{mcc90} state that changes in the brightness of the star by up
to 0.2 magniudes are possible from spots alone.  The variation in the
$V$ magnitude reported since the SXT's return to quiescence has
spanned about 0.17 magnitudes (see Table 1).  \citet{sha94} found that
for spots that are about 750 K below the effective temperature of the
star, the amplitude of the spot modulation is about half as much in
the infrared as in the visual.  They find a 3\% modulation in the
$K$-band for spots similar to those found on RS CVn stars that cover
about 8\% of the star. We also see a 3\% difference in our $K$- band
maxima.  \citet{khr95} applied a model of a spotty Roche lobe filling
star to explain the anomolies in and the long-term variability of the
V616 Mon light curves.  They found that they could explain all of the
peculiarities in the optical and infrared variability of V616 Mon in
its quiescent state by introducing dark spots on the surface of the
K-star with temperatures 10\% - 40\% lower than the mean stellar
temperature.  They successfully modeled the $B$-band light curves
obtained by \citet{mcc86} and \citet{bar91}, even though the light
curves were completely different in appearance.  They implemented one
spot on the hemisphere facing the compact object to explain the light
curve of \citet{mcc86}, and two spots on the opposite hemisphere to
explain the light curve of \citet{bar91}.  \citet{khr95} conclude that
the spotted model is a ``natural explanation of brightness's
inequality in the quadratures of both curves and temporal variations
of the curves' ratio, as well as that of the shape and depth of both
minima.''  The star spots they used reached a few tenths of the
stellar radius, and had temperatures that were 10\% - 40\% lower than
the mean effective stellar temperature. Finally, \citet{bil00} suggested 
that the observed X-ray luminosity of V616 Mon could be attributed to 
coronal activity on the secondary star.  Presumably such coronal 
activity would have associated spots.

Using these dark spot scenarios, we ran WD98 models with a single cool spot
on the surface of the secondary star.  The geometry of the WD98 model places 
$0^{\circ}$ longitude at the line of star centers, measured counterclockwise 
as viewed from the ``north'' (+z) pole.  The angular radius of the spot
corresponds to the angle subtended by the spot radius at the center of
the star.  With $i = 40^{\circ}$, we varied the latitude of the spot from 
$45^{\circ}$ to $135^{\circ}$ ($45^{\circ}$ on either side of the equator), 
the longitude of the spot from $90^{\circ}$ to $162^{\circ}$ (on the leading 
side of the star), the radius of the spot from $60^{\circ}$ (44\% coverage) 
to $9^{\circ}$ (2\% coverage), and the temperature of the spot from 4100
K to 3000 K.  The best fit $\chi^2$ value came from the model of a 3600 K 
spot centered directly on the equator and at a longitude of $135^{\circ}$, 
with a radius of $18^{\circ}$ (4\% coverage).  The model is plotted as a solid 
line in Figure \ref{fig6}. Of course, this same model could be generated 
by several spots, or groups of spots, that have the same average parameters 
as the model presented here.  

An alternative source of phase-dependent contamination is to have
parts of the star brighter than they would otherwise be, either
due to stellar activity (i.e. similar to the faculae on the Sun), or 
possibly due to heating of the star by a disk hot spot. With 
$i = 40^{\circ}$, we varied the hot spot temperature from 4646 K to 4800 K, 
the spot latitude from $45^{\circ}$ to $135^{\circ}$ ($45^{\circ}$ on 
either side of the equator), the spot longitude from $240^{\circ}$ to 
$342^{\circ}$ (on the trailing side of the star), and the spot coverage 
from 100\% (spot radius = $90^{\circ}$) to 30\% (spot radius = 
$50^{\circ}$) of one hemisphere of the star. The best fit model was again 
determined from $\chi^2$ tests.  While several models had very low $\chi^2$ 
values, the model with the lowest value had a 4715 K spot that covered 
100\% of the hemisphere centered directly on the equator and at 
a logitude of 315$^{\rm o}$.  Models with smaller spots ($\ge 60$\% 
coverage) and temperatures varying from 4646 K to 4738 K gave slightly 
higher $\chi^2$ values.  For example, a 4681 K spot centered at the same 
longitude and latitude that covered 64\% of the hemisphere had a $\chi^2$ 
value that was 0.02 higher than that of the best fit model.  

We also used ELC and its genetic code to explore spotted star models.  The
spot parameters are used in the same way as in WD98.  Several populations
were evolved, allowing for either a hot or a cool spot.  Roughly 120,000 
models and fits were computed.  In general, compact hot spots
(with typical angular radii of 10 degrees) on the trailing face of the
star or larger cool spots (with radii of up to 90 degrees) near the leading
face of the star provided good fits.  The cool spot model with the best
$\chi^2$ is shown as the dashed line in Figure \ref{fig6}.  This model has 
$i=40.88^{\circ}$, $q=0.0654$, $a=4.157\,R_{\odot}$.  The 3495~K cool spot 
is centered at a latitude of $172.4^{\circ}$, and a longitude of 
$0.1^{\circ}$, and has an angular radius of $86.8^{\circ}$. The hot spot 
model with the best $\chi^2$ has $i=41.56^{\circ}$, $q=0.0628$, 
$a=4.404\,R_{\odot}$. With a 9050~K spot at a latitude of $83.4^{\circ}$, 
a longitude of $171.1^{\circ}$, and an angular radius of 7.9 degrees.  

Could these hot spots be ``intrinsic'' to the star, or could the star 
be externally heated?  The best fit WD98 model has the entire trailing 
hemisphere of the star slightly hotter than the leading hemisphere.  
This type of temperature distribution may be explained by external 
heating of the secondary star.  One major problem for the external 
heating scenario, however, is that there is little evidence for a {\it luminous} 
hot spot on the accretion disk which could act as the source of heating.  
\citet{mcc00} report a low UV flux.  A blackbody of 12000 K fits the 
dereddened UV points from \citet{mcc00} fairly well, and explains the 
22\% contamination at $B$.  When the blackbody is added to the flux for 
the K4V (open triangles in Figure \ref{fig4}), it nicely accounts for the 
differences between the K4V SED and the observed SED of V616 Mon. Based 
on the total flux of the blackbody and the distance to V616 Mon, the 
radius of the blackbody area is 0.05 $R_{\odot}$.  This amounts to roughly  
2.5\% of the radius of the accretion disk if the disk fills half of its 
predicted Roche lobe.  If this black body is used to represent the hot 
spot on the disk, these results suggest that the hot spot temperature is 
about 12000 K, consistent with hot spot temperatures found for cataclysmic
variable disks \citep{hoa96}.  We also attempted to fit the observed
UV data with a 9000 K blackbody, as found in \citet{mcc95}, however, the
12000 K blackbody better explained the data.  Is this hot spot luminous 
enough to heat the star the required amount to explain the observed light 
curves? According to \citet{bre93}, if the irradiated atmosphere is 
assumed to adopt the structure of a hotter star with an effective temperature
$T\arcmin _{\rm eff}$, then $\sigma T\arcmin$$^4$$_{\rm eff}$ =
$\sigma T^4$$_{\rm eff}$ + $wF_{\rm inc}$ where $F_{\rm inc}$ is the 
flux incident upon the star being heated.  In order to determine the most 
favorable case for a heating scenario, the value of $w$ is taken to be 
1 here (the normal value is approximately 0.5).  To raise the temperature 
of the star from 4600 K to 4715 K, a flux of 
$2.63\times 10^9$ erg s$^{-1}$cm$^{-2}$ is needed.  A 12000 K hot spot 
with a $B$ magnitude that is equal to 22\% of the dereddened $B$-band flux, 
can not be more than 0.86 $R_{\odot}$ ($0.25 R_{L1}$) away from the 
secondary star in order to heat it by 115 K.  To heat the star by 81 K, the 
accretion disk hot spot can not be more than $1.211\, R_{\odot}$ 
($0.35\,R_{L1}$) away from the secondary star.  These solutions require a 
disk radius of 0.65 to $0.75\,R_{L1}$, while the system probably needs a 
much smaller disk in order to allow the accretion stream to travel around 
it to the ``back side'' before impacting the disk.  Unless the phasing of
\citet{lei98} is off by $180^{\circ}$, the heating model seems unlikely.

The best fit ELC hot spot model places a small 9050 K spot near the 
equator on the end of the star opposite from the L$_1$ point.  Since it is 
unlikely that such a spot could be a result of external heating, we explore 
the possibility of faculae on the secondary star.  Solar faculae are found 
around groups of sunspots and are normally visible near the limb of a star.  
The lifetime of an average solar faculae is about 15 days, while that of a 
large faculae which dominates solar variations can last almost three months 
\citep{cha77}.  Since individual star spots can exist for much longer 
timescales than their solar counterparts, it may be possible for stellar 
faculae to also exhibit much longer lifetimes.  However, even though it may 
be possible for faculae to remain stable over the baseline of our 
observations, they do not appear to be able to account for a spot that has 
a temperature of almost twice that of the effective temperature of the star.  
\citet{cha77} find that in general, the facular contrast has a $\lambda^{-1}$ 
dependence.``White light'' faculae on the sun are associated with bright 
regions, but in the infrared, they appear darker than the quiet photosphere 
\citep{fou89}. It does not appear as though faculae on the secondary star in 
V616 Mon could have parameters consistent with the ELC hot spot.

\subsection{The Adopted Model and its Uncertainties}

Given the plausibility of the physical nature of the dark spots, and their
ability to reproduce light curves of vastly differing shapes through
migration, we suggest that cool spots on the surface of the secondary
star are the cause of the phase dependent deviations in the
ellipsoidal variations seen in the $J$-, $H$-, and $K$- band light
curves.  From the SED shown in Figure \ref{fig4}, we adopt a model that
has no light from an accretion disk.

The best fitting inclination from the WD98 models is $i=40^{\circ}$
and from the ELC models is $i=41.5^{\circ}$.
Based on errors in the amount of infrared diluting light in the system, 
and the ability of the $\chi^2$ tests to distinguish between model fits 
to the observed data, the uncertainty in the WD98 value is about 1.5 
degrees. The ELC code and its genetic routine provide an easy way to estimate 
the uncertainties on the fitted parameters, especially the inclination.
We simply plotted the $\chi^2$ values as a function of the inclination
and looked to see how the value of $\chi^2$ changes near the minimum.
Figure \ref{fig7} shows the ``lower envelope'' of the $\chi^2$ values as a 
function the inclination, which is basically the smallest $\chi^2$ value 
within each $0.1^{\circ}$ bin.  The $\chi^2$ values were scaled to give 
$\chi^2_{\rm nu}=1$ at the minimum.  Although not all bins are filled,
it appears that the {\em formal} statistical uncertainty in the inclination 
is rather small, on the order of $2^{\rm \circ}$ or so.  We estimate that 
the systematic error caused by contamination of disk light at the few 
percent level is on the order of $1^{\circ}$.  We adopt as our final 
inclination $i=40.75\pm 3^{\circ}$.

Our derived orbital inclination angle is inconsistent with HRHSA's,
but agrees with SNC's inclination determined from their $K$-band
light curve of V616 Mon.  We do not see any evidence for an accretion
disk hot spot affecting the infrared light curves, and thus our
results differ from those of FR.

\section{Discussion}

With the excellent match of the observed photometry with a K4V SED, we
find that the optical and infrared contamination from a constant
source of light in the system is $\le$3\%, with an exception at $B$.
Combining HST UV observations (assuming a 12000 K blackbody for the
hot spot on the accretion disk), we can explain the observed SED from
the ultraviolet to the infrared.  We first ran models for an
uncontaminated K4V star and found a best fit orbital inclination of
40$^{\rm o}$.  This model could not fit the unequal maxima of the
infrared light curves, as a purely ellipsoidal model exhibits equal
maxima.  In order to determine the phase dependent nature of the
ellipsoidal deviations in the V616 Mon light curves, we have
considered three possibilities.  We ruled out models for a migrating
hot-spot on the accretion disk based on the low level of optical and 
infrared diluting light in the system.  However, models with either cool
spots or hot spots on the surface of the star fit the data nearly equally 
as well, although the heated spot model was slightly better.  However, 
each model has its own problems.

There are several main problems with the heated face models.  Firstly,
the heated part of the star is the trailing hemisphere.  This means
that the hot spot needed to heat it would have to be on the opposite
side of the accretion disk than is normally expected.  Secondly, the
heating mechanism must be uncomfortably close to the secondary star in
order to heat it and explain the data.  There does not appear to be
sufficient luminosity in the system to heat the star from the
suggested distance of a hot spot on the accretion disk ($0.5\,R_{L1}$).
This scenario might work if the phasing were to be off by $180^{\circ}$
(but this possibility seems unlikely).

Contrary to the heated models, the dark spot models seemed to be
physically viable, but even these models had their own peculiarities.
The best fit WD98 dark spot models placed the spot directly on or very 
close to the equator. Also, the model had one spot on the leading face 
of the secondary star.  Thus this model may require a special scenario, 
however it is likely that this same model could be generated by multiple 
spots that have the same average parameters. Nonetheless, this model is 
consistent with both the long-term as well as current observations of 
V616 Mon.  Spots are presumed to be a natural occurrence on the surfaces 
of late-type stars.  Any such spots, however, must be nearly unchanging 
on multi-year timescales. Dark spots have also been used to explain the 
changing light curves of Cen X-4 \citep{mcc90} as well as other binary 
systems.  Detailed long-term monitoring of V616 Mon and other SXTs should 
prove extremely useful in determining the behavior of spots (if present) 
on long timescales.

With the orbital inclination angle determined as $i = 40.75\pm 3^{\circ}$, 
we can estimate the mass of the primary and other interesting parameters.
The other observed quantities that are needed for this exercise are the
(projected) radial velocity and rotational velocity of the secondary star 
($K_2 = 433 \pm 3$ km s$^{-1}$ and $v\sin i=83\pm 5$ km s$^{-1}$, MRW),
and the orbital period ($P=0.323016$ days, McClintock \& Remillard 1986).
The ELC code was used to numerically find the relationship between
$K_2$, $v\sin i$, and the mass ratio $q$, and a simple Monte Carlo code was
used to compute the uncertainties on the binary system parameters,
assuming the quoted uncertainties on all of the input values above
are $1\sigma$.  We find a black hole mass of $11.0 \pm 1.9 \,M_{\odot}$
and a mass of $0.68 \pm 0.18\,M_{\odot}$ for the secondary star.  
Note that this mass is consistent with that of a ``normal'' K4V \citep{gra92}.

The distance to the source depends on its luminosity and extinction. We 
used the synthetic photometry computed 
from the {\sc NextGen} models\footnote{\tt
ftp://calvin.physast.uga.edu/pub/NextGen/Colors/} to compute the expected
absolute $J$, $H$, and $K$ magnitudes of the star from its
temperature, radius, and surface gravity.  For this exercise we
adopt a $1\sigma$ error on the effective temperature of 200~K, which
is roughly one subclass in K \citep{gra92}.  We used the color excess of
$E(B-V)=0.39\pm 0.02$ given by \citet{wu76}, and extinction law
of \citet{car89} to find the extinctions in the three bands, assuming
$A_V=3.1E(B-V)$.   Using our infrared photometry (Table 1) and the quoted 
$1\sigma$ errors, we used another simple Monte Carlo procedure to compute 
the absolute magnitude and the distance modulus and their uncertainties
in each of the $J$, $H$, and $K$ filters.  The results are summarized in 
Table 4.  Our adopted value of the distance is the simple average of the 
distances derived for the three filters: $d=1164\pm 114$ pc.  This distance 
is consistent with that found by SNC ($660 \le d\le 1450$ pc with
the most probable value at 1050 pc).

In order to understand the outbursts of SXTs, we need to be able to
model them with refined input parameters.  As discussed in Paper 1,
the orbital inclination angle can play an important role in the
appearance of an SXT outburst. \citet{esi00} used a black hole mass of
$4.5\,M_{\odot}$, an orbital inclination angle of $i=65^{\circ}$, and a
distance of 1.4 kpc as inputs to their advection dominated accretion
flow (ADAF) models.  While the distance they used was close to the one
we find here, the black hole mass and inclination angle are not.  
Any new models that are constructed to explain the outburst of V616 Mon
should take into account our results for the mass of the black
hole, the orbital inclination angle, and the distance.

\begin{figure}
\figurenum{4}
\plotone{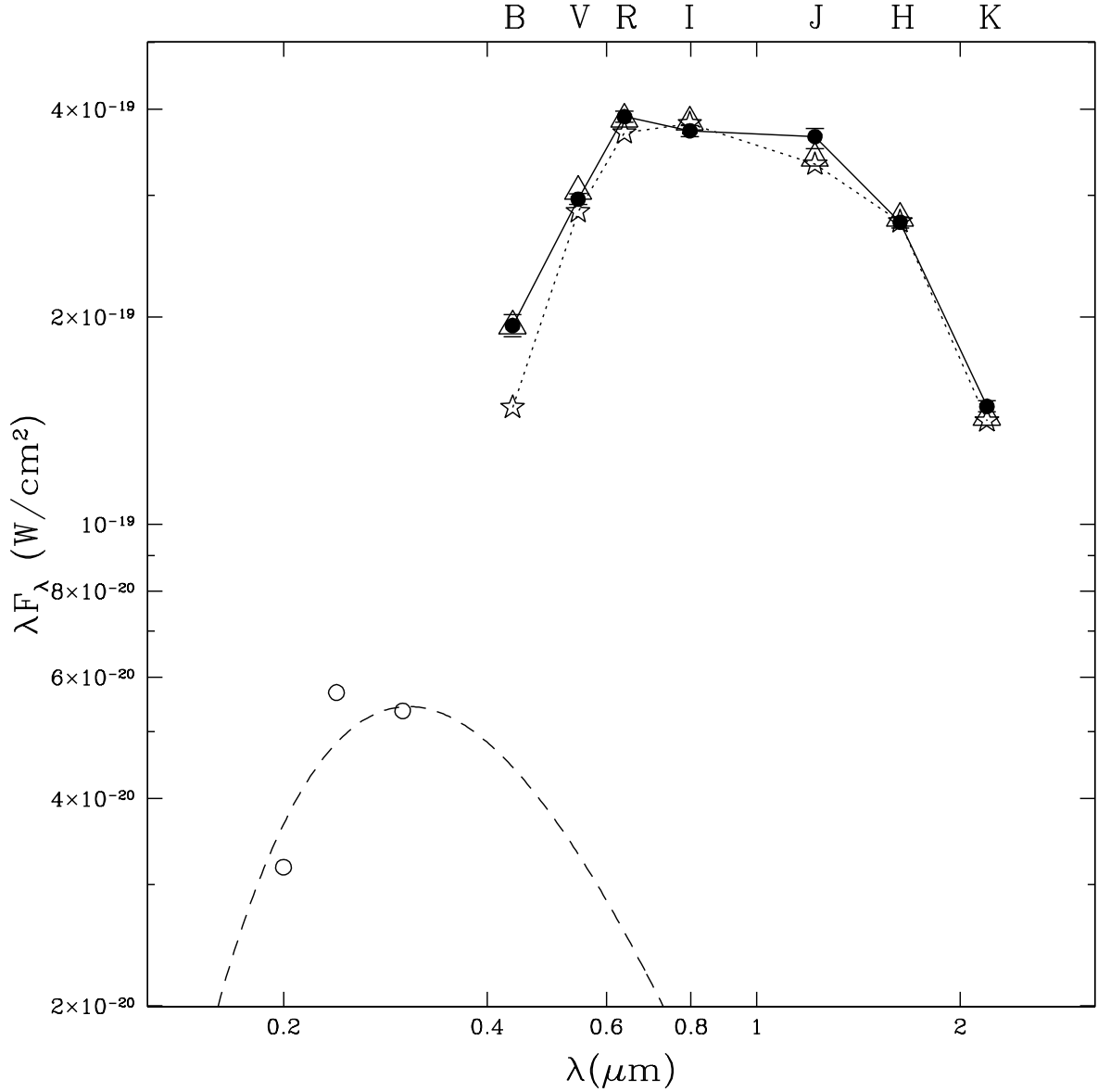}
\caption{The 2000/2001 V616 Mon optical/infrared spectral energy 
distribution (SED) dereddened by A$_{\rm V}$ = 1.21 magnitudes 
(filled circles), compared to the SED of a K4V (open stars) normalized at 
$H$.  The two SEDs match well within the errors with small exceptions at 
$B$ and $J$. The dashed line is a 12000 K blackbody representing the hot 
spot on the accretion disk.  This blackbody is consistent with the dereddened
ultraviolet observations taken by \citet{mcc00} (open circles), and
when added to the SED of the K4V (open triangles), explains the
observed excess at $B$, and provides an excellent fit at the other 
wavelengths. \label{fig4}}
\end{figure}

\begin{figure}
\figurenum{5}
\plotone{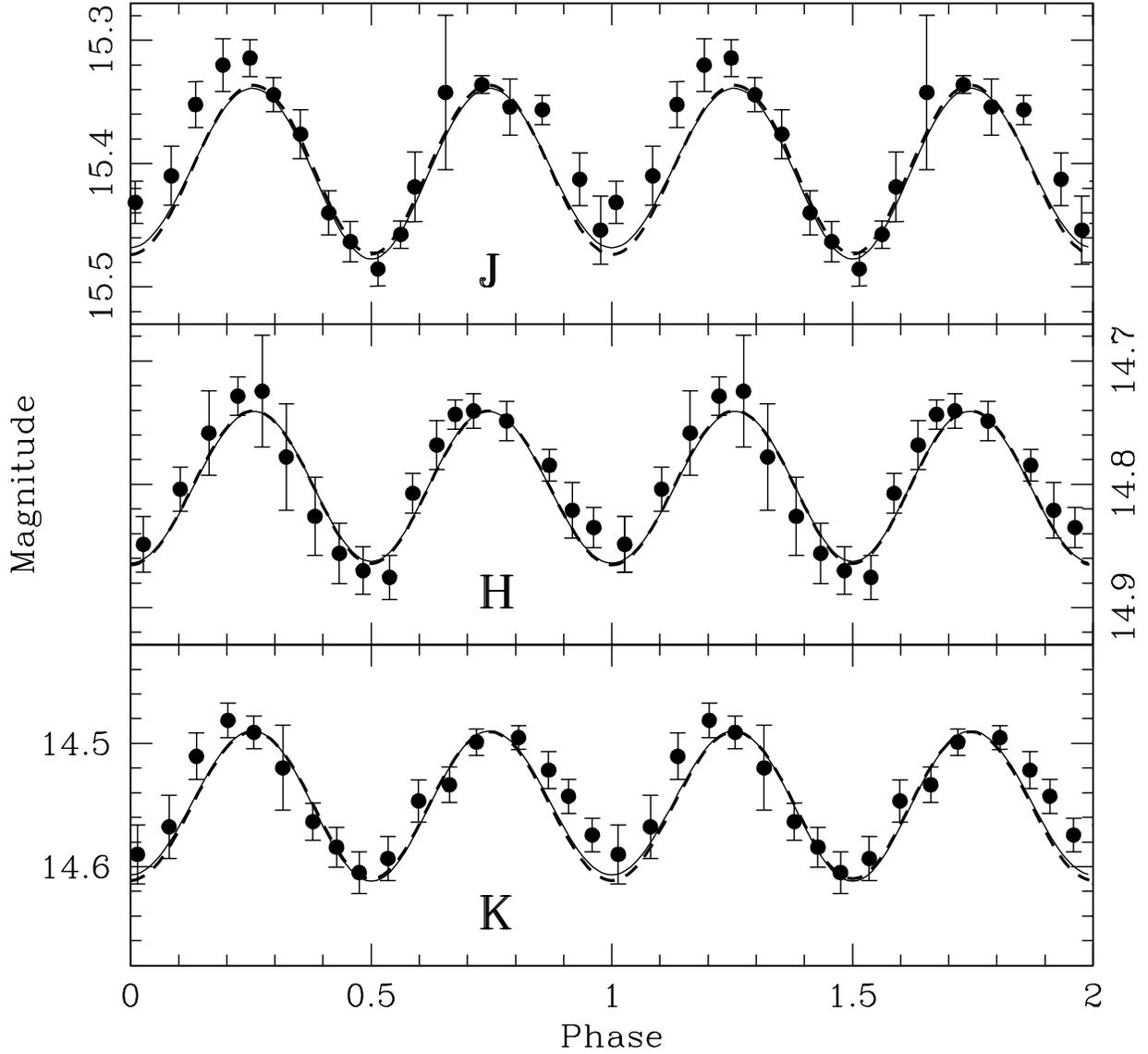}
\caption{The data from Figure \ref{fig2} (points) and the best fitting 
models for a K4V secondary star.  The WD98 model (solid line) was 
determined from $\chi^2$ tests and has $i$ = 40$^{\circ}$. The ELC model
(dashed line) was found using its grid search program, and also has 
$i$ = 40$^{\circ}$.  These simple ellipsoidal models are nearly 
indistinguishable. A small excess is present in the rise to the first 
maximum, while the second minimum is more shallow than either model. 
\label{fig5}}
\end{figure}

\begin{figure}
\figurenum{6}
\plotone{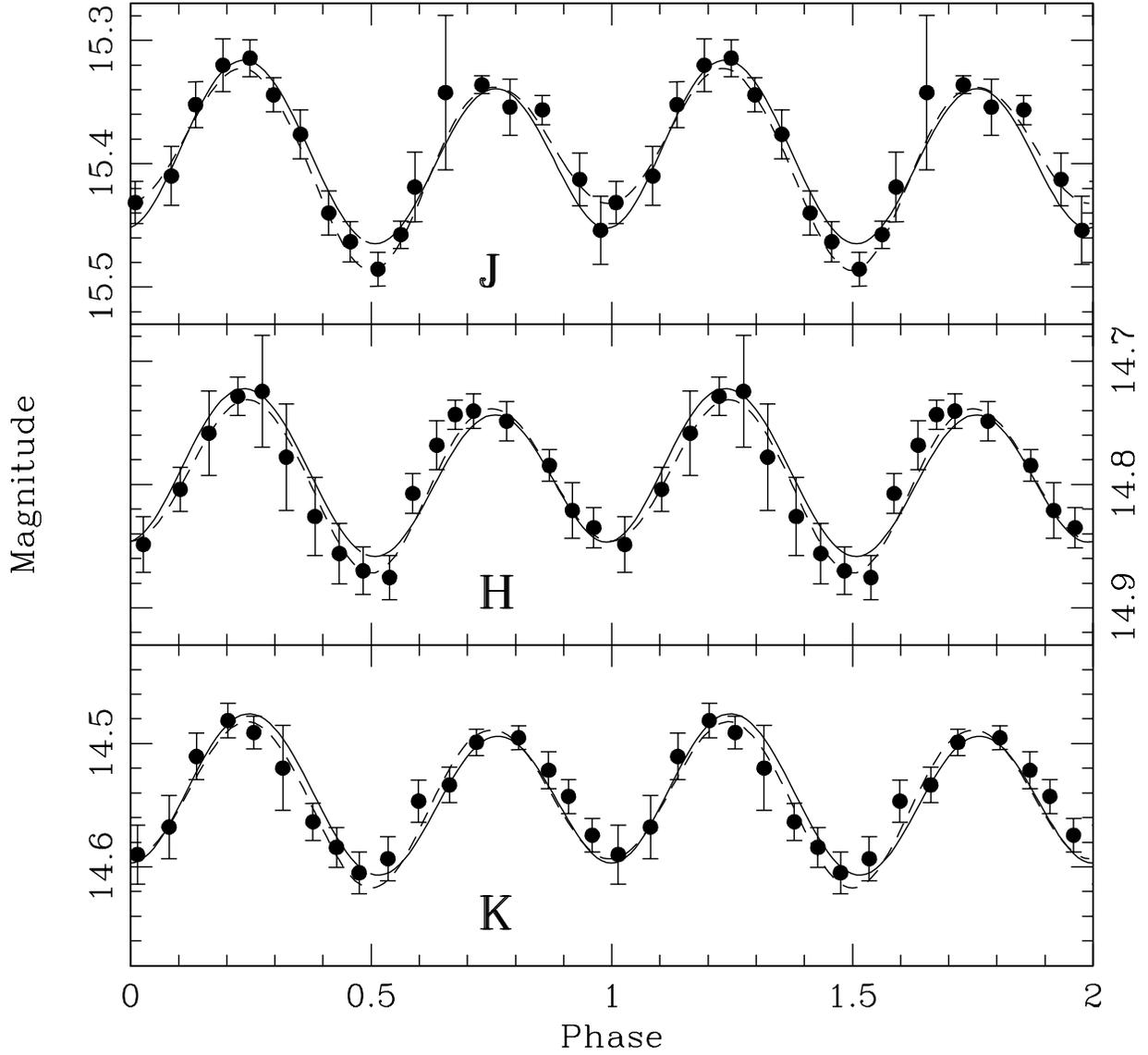}
\caption{The data from Figure \ref{fig2} (points) and the best fitting 
cool spot models.  The solid line is the WD98 model ($i$ = 40$^\circ$) 
with a 3600 K spot centered on the equator and at a longitude of 
135$^{\circ}$ with a radius of 18$^{\circ}$ (4\% coverage).  The dashed 
line is the ELC model ($i$ = 41$^\circ$) with a 3495 K spot centered at
a latitude of 172$^{\circ}$ and a longitude of 0$^{\circ}$ with a radius 
of 87$^{\circ}$ (93\% coverage). The hot spot models had similar fits, 
however, they could not be easily explained. \label{fig6}}
\end{figure}

\begin{figure}
\figurenum{7}
\plotone{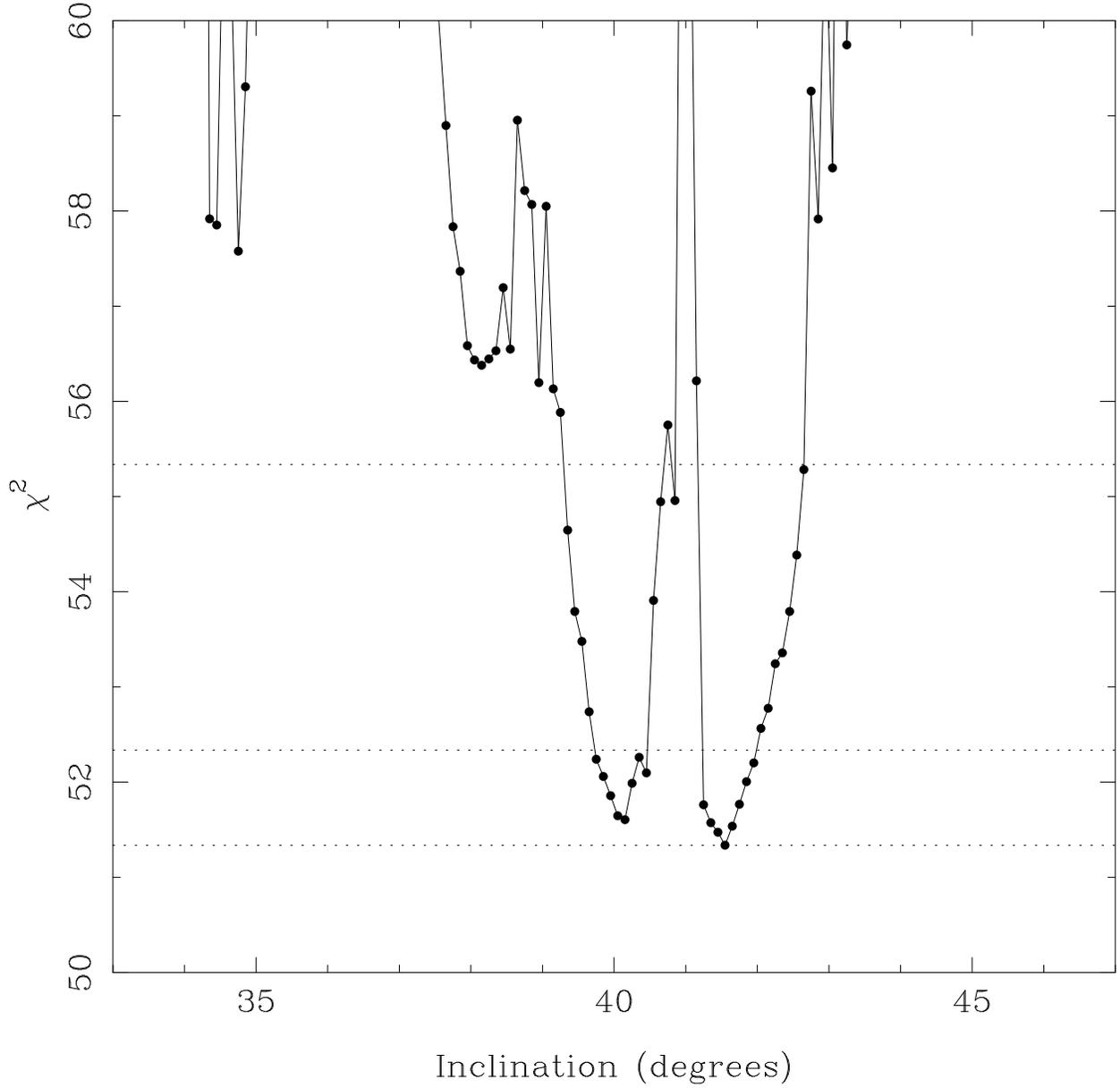}
\caption{The ``lower envelope'' of the
$\chi^2$ values as a function of the inclination found by the ELC genetic
fitting routine for the spotted secondary star models (with no disk).  
The dotted lined denote the minimum $\chi^2$ value $\chi^2_{\rm min}$, 
$\chi^2_{\rm min}+1$, and $\chi^2_{\rm min}+4$. \label{fig7}}
\end{figure}

We assumed that the accretion disk contributes essentially no light in
the infrared.  Based on this assumption, we adopted a value of the 
inclination of $i=40.75^{\circ}$.   As the SED in Figure 
\ref{fig4} shows, there 
are good reasons to think that this assumption is valid.  Given how 
critical this assumption is (i.e.\ a disk contamination of 50\% would 
drive the inclination up to around $60^{\circ}$), it would be extremely 
worthwhile to test it in as many ways as possible. \citet{sha99} obtained 
a $K$-band spectrum of the source with modest resolution (47~\AA\ FWHM) 
and signal-to-noise ($\approx 30$) and concluded that the $2\sigma$ upper 
limit on the accretion disk contamination is $\le 27\%$ in the $K$-band.  
FR, however, critized this result on the grounds that template stars with 
inappropriate surface gravities were used and that an inappropriate 
spectral feature (the $\lambda 2.29\mu\,{\rm m}$$~^{12}$CO(2,0) bandhead) 
was used.  It should be relatively easy to obtain a much better infrared 
spectrum using the latest generation of infrared spectrographs available 
on the 8 to 10 meter class telescopes.  One could observe a wider variety
of templates, and also make use of the increasingly detailed spectral
models for cool stars (for example the {\sc NextGen} models discussed in 
Leggett et al.\ 2001) to derive a more definitive upper limit on the disk 
contamination in the infrared.

The shape of the X-ray outburst light curves of V616 Mon and GU Mus
are very similar [as can be seen in Figure 1 of \citet{kin96}] with
one main difference, the time of their secondary maxima.  The light
curve of V616 Mon began the rise to its second X-ray maximum about 40
days after its initial outburst, while the light curve of GU Mus shows
this rise beginning about 60 days afterward.  Both secondary maxima
lasted for about 30 days.  Based on the orbital period and radial
velocity of the secondary star, the GU Mus system (P = 10.38 hr; 
$K_2 = 406 \pm 7$ km s$^{-1}$) is larger than the V616 Mon system 
(P = 7.75 hr; $K_2 = 433 \pm 3$ km s$^{-1}$).  V616 Mon has a larger 
primary mass ($M_{1,\rm Mon}$ = 11.0, $M_{1, \rm Mus}$ = 6.95), and we 
see it at more of a face-on angle ($i_{\rm Mon}$ = 40.75$^\circ$, 
$i_{\rm Mus}$ = 54$^\circ$) than GU Mus.  Could the larger size of the 
system account for the increase in time between X-ray outburst maxima, 
or could it be related to the mass of the black hole?  Close study of 
more SXTs is needed before this question can be answered with any certainty.  

In Paper 1, based on the comparison of optical and infrared colors, we
determined a spectral type of K4 for the secondary star in the GU Mus
system.  The colors were intermediate to those of a K4 dwarf and giant.
As discussed above, we also determine a spectral type of K4 for V616 Mon.
In this case, however, the colors are consistent with those of a K4
dwarf.  By Kepler's Third Law, the ratio of the secondary stars' Roche lobe
radii is simply related to the ratio of the orbital periods:
$R_{\rm Mon}/R_{\rm Mus}=(P_{\rm Mon}/P_{\rm Mus})^{2/3}=0.82$.
A K4V star nominally has a radius of $0.70\,R_{\odot}$ on the zero-age
main sequence \cite{gra92}.  Such a star would nearly fill the Roche lobe 
in V616 Mon ($R_2=0.80\pm 0.07\,R_{\odot}$), whereas it would only fill 
$\approx 80$\% of the Roche lobe in GU Mus.  Surprisingly, accurate
multi-wavelength photometry appears to be somewhat sensitive to these
small radius differences.

The nature of the mass distribution of stellar-mass black holes has been 
in question for many years now.  Based on observational evidence for seven 
stellar black holes with low-mass companions, \citet{bai98} attempted to 
constrain this distribution.  Using Bayesian statistics, they suggested 
that six of the seven systems had black hole masses clustered around 
7 $M_{\odot}$.  The other system, V404 Cyg, was thought to be drawn from 
a different distribution.  Models of the evolution and supernova explosions 
of massive stars should be able to predict the mass distribution of 
stellar-mass black holes.  However, because of their findings, \citet{bai98} 
suggested considering a new underlying mechanism for the origin of the black 
hole mass distribution in low-mass X-ray binaries.

The mass we found for GU Mus in Paper 1 ($M_1 = 6.95\pm0.6 M_{\odot}$) 
is consistent with, but more precise, than the mass adopted by 
\citet{bai98}.  In the case of V616 Mon, \citet{bai98} adopted a wide 
range of inclinations reported in the literature 
($31\le i\le 70.5^{\circ}$).  As a result, the allowed range of the mass 
of the black hole was rather large. The more precise mass we detemine 
for V616 Mon is well above the cluster of masses near $7\,M_{\odot}$ and
falls nicely in the mass range plotted in their Figure 1 for V404 Cyg.  
Although the orbital period and evolutionary status of the secondary star 
in V404 Cyg are significantly different from V616 Mon, their similar black 
hole masses may say something about the amount of post-supernova mass 
transfer that has taken place in both these systems.  The sample size of 
SXTs with measured mass functions has doubled since the analysis of 
\citet{bai98} was done, so it would be worthwhile to revisit the issue 
of the distribution of stellar black hole masses.  Indeed, the four most 
recently determined SXT optical mass functions are all rather large:  
$6.0\pm 0.36\,M_{\odot}$ for XTE J1118+480 \citep{mcc01}, 
$6.86\pm 0.71\,M_{\odot}$ for XTE J1550-564 \citep{oro01}, 
$9.5\pm 3\,M_{\odot}$ for GRS 1915+105 \citep{gre01}, and 
$7.4\pm 1.1\,M_{\odot}$ for XTE J1859+226 \citep{fil01}.  It is likely 
that some of these systems will have black hole masses that are also 
well above the cluster of masses near $7\,M_{\odot}$.
In addition, better measurements for more of the known systems
can be used to reduce the uncertainties in the mass distribution.

\acknowledgments

This research was supported by a Grant-in-Aid of Research from the National 
Academy of Sciences, through Sigma Xi, The Scientific Research Society. DMG 
holds an American fellowship from the American Association of University 
Women Educational Foundation.  TEH wishes to acknowledge support from NASA 
through grant number GO-8618 from the Space Telescope Science Institute, 
which is operated by the Association of Universities for Research in 
Astronomy under NASA contract NAS 5-26555. This research has made use of 
the SIMBAD database, operated at CDS, Strasbourg, France, and NASA's 
Astrophysics Data System Abstract Service.

\clearpage

\clearpage
 
\begin{deluxetable}{ccccccccc}
\tabletypesize{\scriptsize}
\tablewidth{0pt}
\tablenum{1}
\tablecaption{Quiescent Infrared and Optical Colors of V616 Mon}
\tablehead{
\colhead{Reference} &\colhead{V} &\colhead{B - V} &\colhead{V - R}
&\colhead{V - I} &\colhead{J} &\colhead{J - H} &\colhead{J - K} &\colhead{Year\tablenotemark{a}}
}
\startdata
1 & 18.35 & 1.35 & ... & ... & ... & ... & ... & 1976 \\
2 & 18.2$\pm$0.1 & ... & ... & ... & ... & ... & ... & 1978 \\
3 & 18.22$\pm$0.03 & 1.25 & 1.08 & ... & ... & ... & ... & 1986/1987 \\
4 & ... & ... & ... & ... & 15.44$\pm$0.01 & ... & 0.89 & 1990 \\
5 & 18.20$\pm$0.05 & ... & ... & ... & ... & ... & ... & 1992 \\
5 & 18.25$\pm$0.08 & ... & ... & ... & ... & ... & ... & 1995/1996 \\
6 & ... & ... & ... & ... & 15.6$\pm$0.1 & 0.76 & ... & 1996 \\
5 & 18.37$\pm$0.05 & ... & ... & ... & ... & ... & ... & 1998 \\
7 & 18.27$\pm$0.04 & 1.21$\pm$0.04 & 0.83$\pm$0.04 & 1.57$\pm$0.04 & 15.40$\pm$0.04 & 0.60$\pm$0.04 & 0.85$\pm$0.04 & 2000/2001 \\
&&&&&&& \\
K4V (A$_{\rm V}$ = 1.21 mag) & & 1.46 & 0.82 & 1.65 & & 0.70 & 0.88 \\
K4III (A$_{\rm V}$ = 1.21 mag) & & 1.80 & 0.93 & 1.99 & & 0.86 & 1.08 \\
\enddata
\tablenotetext{a}{Year the data were taken}
\tablerefs{(1) Oke 1977; (2) \citet{mur80}; (3) HRHSA; (4) SNC; (5) McClintock \& Remillard 2000; (6) FR; (7) This Paper}
\end{deluxetable}{}


\begin{deluxetable}{lc}
\tablenum{2}
\tablecaption{Wavelength Independent WD98 Input Parameters for V616 Mon}
\tablehead{
\colhead{Parameter}
 &\colhead{Value}
}
\startdata
Orbital Period (days) & 0.323016 \\
Ephemeris (HJD phase 0.0)\tablenotemark{a} & 2450000.025 \\
Semi-Major Axis ($R_{\odot}$) & 4.6 \\
Orbital Eccentricity & 0.0 \\
Temperature of K4V Secondary (K) & 4600 \\                 
Mass Ratio (M$_2$/M$_1$) & 0.067 \\
Atmosphere Model &  Kurucz \\
Limb Darkening Law & Square-root \\ 
Secondary Star Gravity Darkening Exponent & $\beta$=0.40 \\ 
Secondary Star Bolometric Albedo\tablenotemark{b} & 0.676 \\
\enddata
\tablenotetext{a} {From Leibowitz, Hemar, \& Orio (1998); Corresponds to phase 0.0 for light curves presented in this paper, and phase 0.5 from their paper.}
\tablenotetext{b} {From \citet{nor90}}
\end{deluxetable}{}


\begin{deluxetable}{cccc}
\tablenum{3}
\tablecaption{Wavelength Dependent WD98 Input Parameters for V616 Mon}
\tablehead{
\colhead{Secondary Star Parameter}
 &\colhead{J}
&\colhead{H}
 &\colhead{K}
}
\startdata
Monochromatic Luminosity ($L_{\odot}$) & 0.223 & 0.281 & 0.296 \\
Square-root Limb Darkening Coefficient x$_\lambda$ & 0.110 & -0.063 & -0.166 \\
Square-root Limb Darkening Coefficient y$_\lambda$ & 0.531 & 0.652 & 0.724 \\
\enddata
\end{deluxetable}{}


\begin{deluxetable}{lr}
\tablenum{4}
\tablecaption{Derived Parameters for V616 Mon}
\tablehead{
\colhead{Parameter}
&\colhead{Value}}
\startdata
Black Hole Mass $M_1$ ($M_{\odot}$) & $11.0\pm 1.9$ \\
Secondary Star Mass $M_2$ ($M_{\odot}$) & $0.68\pm 0.18$ \\
Total Mass ($M_{\odot}$) & $11.70\pm 2.05$ \\
Orbital Separation $a$ ($R_{\odot}$) & $4.47\pm 0.27$ \\
Secondary Star Radius $R_2$ ($R_{\odot}$) & $0.80\pm 0.07$ \\
Secondary Star Gravity ($\log g$ cgs) & $4.46\pm 0.04$ \\
Absolute Magnitude, $J$  & $4.77\pm 0.23$ \\
Absolute Magnitude, $H$  & $4.17\pm 0.20$ \\
Absolute Magnitude, $K$  & $4.11\pm 0.20$ \\
Distance Modulus, $J$ (mag) & $10.29\pm 0.23$ \\
Distance Modulus, $H$ (mag) & $10.40\pm 0.21$ \\
Distance Modulus, $K$ (mag) & $10.30\pm 0.21$ \\
Adopted Distance (pc) & $1164\pm 114$ 
\enddata
\end{deluxetable}{}

\end{document}